\documentclass[5p,twocolumn]{elsarticle}

\usepackage{graphicx}
\usepackage{dcolumn}
\usepackage{pifont}
\usepackage{bm}
\usepackage{multirow}
\usepackage{amsmath}
\usepackage{float}
\usepackage{txfonts}
\usepackage[version=3]{mhchem}

\usepackage[usenames,dvipsnames]{xcolor}
\definecolor{blue}{RGB}{0,0,225}
\definecolor{cream}{RGB}{222,217,201}
\definecolor{red}{RGB}{225,0,0}

\journal{arXiv}

\begin{document}
\title{Manifestation of the thermoelectric properties in Ge-based halide perovskites}

\author[kimuniv-m,kimuniv-n]{Un-Gi Jong\corref{cor}}
\ead{ug.jong@ryongnamsan.edu.kp}
\author[kimuniv-m]{Chol-Jun Yu\corref{cor}}
\cortext[cor]{Corresponding author}
\ead{cj.yu@ryongnamsan.edu.kp}
\author[kimuniv-m,kimuniv-n]{Yun-Hyok Kye}
\author[kimuniv-m]{Song-Nam Hong}
\author[kimuniv-m]{Hyon-Gyong Kim}

\address[kimuniv-m]{Computational Materials Design (CMD), Faculty of Materials Science, Kim Il Sung University, Ryongnam-Dong, Taesong District, Pyongyang, Democratic People's Republic of Korea}
\address[kimuniv-n]{Natural Science Center, Kim Il Sung University, Ryongnam-Dong, Taesong District, Pyongyang, Democratic People's Republic of Korea}

\begin{abstract}
In spite of intensive studies on the chalcogenides as conventional thermoelectrics, it remains a challenge to find a proper material with high electrical but low thermal conductivities.
In this work, we introduced a new class of thermoelectrics, Ge-based inorganic halide perovskites \ce{CsGeX3} (X = I, Br, Cl), which were already known as a promising candidate for photovoltaic applications.
By performing the lattice-dynamics calculations and solving the Boltzmann transport equation, we revealed that these perovskites have ultralow thermal conductivities below 0.18 W m$^{-1}$ K$^{-1}$ while very high carrier mobilities above 860 cm$^2$ V$^{-1}$ s$^{-1}$, being much superior to the conventional thermoelectrics of chalcogenides.
These results highlight the way of searching high-performance and low-cost thermoelectrics based on inorganic halide perovskites.
\end{abstract}

\maketitle

\section{Introduction}
Thermoelectrics, which are functional materials that convert a small amount of losing waste heat into electricity, have attracted a renewed interest in the drive for ``green" energy~\cite{Snyder08}.
Much of the recent work on thermoelectrics has focused on binary chalcogenides such as selenides (SnSe, PbSe, CuSe$_2$)~\cite{Zhao14,Nishimura,Aseginolaza,Skelton16prl,Zhao16,Liu12,Voneshen} and tellurides (GeTe, SnTe, PbTe)~\cite{Wu14,Xia180,Pei11,Xia181,Tan15}.
These have a fairly high figure of merit for thermoelectric performance, which is defined as $ZT=S^2\sigma T/(\kappa_e+\kappa_l)$ with the Seebeck coefficient and electrical conductivity ($S$ and $\sigma$), and the electronic and lattice thermal conductivities ($\kappa_e$ and $\kappa_l$)~\cite{Xia182,He16,Lu15,Pei110}.
In particular, Sb/In-codoped $\beta$-GeTe~\cite{Hong18} and S-doped PbTe~\cite{Wu141} were shown to have a $ZT$ over 2 due to their ultralow lattice thermal conductivity at a high temperature range over 700 K in combination with favorable electrical properties.
However, these chalcogenides include the constituent Se or Te, which are relatively scarce in the earth's crust, causing a high cost of device.

%Some relevant thermoelectrics
For practical applications of thermoelectrics, their $ZT$ is required to be above 1 at the given temperature, which can be achieved by reducing the thermal conductivity while keeping a high thermopower factor ($S^2\sigma$)~\cite{Skelton16prl}.
However, it is not easy to find such a material with high electrical but low thermal conductivities.
Although low-dimensional materials~\cite{JuChen,Nguyen,Maurie} and nano-structured composites~\cite{Miura} have shown to possess a high value of $ZT$, their production in large-scale can be realized with much higher cost compared to bulk materials.
In this context, searching a new type of bulk thermoelectrics with a high $ZT$ and low synthesis cost, which should neither contain Se and Te nor be in the form of nano-phase, is a desirable route to the development of practical thermoelectrics.
Some oxides~\cite{Ito} and phosphides~\cite{Shen18} with stereochemically active lone pair electrons have been lately proposed as such a promising candidate for high performance thermoelectrics, but the highest value of their $ZT$ was found to be below 0.7, being lower than the required value.

%Halide perovskites
Halide perovskites have recently proved to be not only excellent optoelectronic materials but also very promising thermoelectrics.
They have ultralow thermal conductivities, which are thought to be related with the strongly anharmonic lattice dynamics observed in the cubic phase~\cite{Jong19prb,Marronnier17jpcl,Yang17jpcl,Zhao17apl}.
For instance, the organic-inorganic hybrid iodide perovskite \ce{CH3NH3PbI3} was shown to have a ultralow thermal conductivity of 0.5 and even 0.05 W m$^{-1}$ K$^{-1}$ by experiment~\cite{Pisoni14,Qian16} and first-principles simulation~\cite{Whalley16prb}, respectively.
What is more, the halide perovskites possess a high carrier mobility~\cite{Wang15pccp,Oga14}, which is essential for photovoltaic applications and allows good electrical conductance.
In experiment, the all-inorganic halide perovskites, such as $\delta$-\ce{CsPbI3} and $\gamma$-\ce{CsSnI3}, were confirmed to have a high electrical conductivity (e.g. 282 S cm$^{-1}$ in $\gamma$-CsSnI$_3$) and high hole mobility (394 cm$^2$ V$^{-1}$ s$^{-1}$) with a very low lattice thermal conductivity (0.38 W m$^{-1}$ K$^{-1}$)~\cite{Lee17pnas,Kovalsky17jpcc}.
Together with the low synthesis cost, these highlight the potential application of the halide perovskites to heat-to-electricity conversion in high performance and practical thermoelectric devices.

Previous theoretical and experimental studies have shown that the all-inorganic Ge-based halide perovskites are a promising optoelectronic material for photovoltaic applications~\cite{Jong19ic,Huang16prb,Thirumal15jmca,Walters18jpcl}.
%In this work, we have performed first-principles calculations to investigate vibrational and thermal properties of the Ge-based halide perovskites \ce{CsGeX3} (X = I, Br, Cl), with a particular focus on revealing the relation between anharmonic soft-mode instability and ultralow lattice thermal conductivity, and on developing a practicable procedure to the theoretical challenge of modeling transport properties.
In this work, we have performed first-principles calculations to investigate lattice vibrational and electronic transport properties of the Ge-based halide perovskites \ce{CsGeX3} (X = I, Br, Cl), with a particular focus on revealing the ultralow lattice thermal conductivity and high carrier mobility which are key factors for enhancing the thermoelectric figure of merit ZT in the thermoelectric materials.

\section{Methods}
All density functional theory (DFT) calculations were performed by using the projector augmented wave (PAW)~\cite{paw1,paw2} method as implemented in the Vienna ab initio simulation (VASP) package~\cite{vasp1,vasp2}. The PBEsol~\cite{PBEsol} functional was adopted to describe the exchange-correlation interaction between the valence electrons. We set the kinetic energy cutoff as 800 eV for the plane-wave basis and the $k$-point mesh as 8$\times$8$\times$8 for the Brillouin zone integration. With such parameters, we perform the self-consistent calculations to get the total energies with tight convergence threshold of 10$^{-8}$ eV and optimize the unit cells until forces on the atoms reach below 1 meV/$\AA$, which were proved to be sufficient for the DFT total energy and phonon energy convergences within 0.01 meV per unit cell.
The valence electronic configurations of the atoms given in the provided PAW bases are Cs--5s$^2$5p$^6$6s$^1$, Cl--3s$^2$3p$^5$, Br--4s$^2$4p$^5$, I--5s$^2$5p$^5$, and Ge--4s$^2$4p$^2$.
In this work, we did not repeat the structural optimizations of the unit cells for the inorganic perovskite \ce{CsGeX3} (X = I, Br, Cl) in the rhombohedral phase with $R3m$ space group, but used the previously determined values in our previous work~\cite{jongIO}.
Likewise, we utilized the effective masses of electron and hole, and electronic band structures, calculated considering spin-orbit coupling (SOC) effects in our previous work~\cite{jongIO}.

Lattice dynamics calculations were carried out by using the finite displacement method, as implemented in the Phonopy code~\cite{phonopy}.
We used the VASP code as a calculator to obtain forces on atoms in displaced supercells generated by the Phonopy package.
We calculated the vibrational properties including phonon dispersions and phonon density of states (DOS), using the 2$\times$2$\times$2 supercells and reduced the $k$-point sampling for force calculations in accordance to larger size of the supercell.
The Phonopy package provided 8 displaced supercell structures.
For these supercells, the 30$\times$30$\times$30 mesh for the $q$ points was used in the calculation of the phonon DOS.
Non-analytic correction of phonon dispersion at $\Gamma$ point was performed using the mixed-space approach with Born effective charges and macroscopic static dielectric constant computed by density functional perturbation theory (DFPT)~\cite{jongIO}.
The lattice thermal conductivity was calculated considering up to the three-phonon scattering as implemented in the Phono3py package~\cite{phono3py}.
Describing the three-phonon interactions requires a significantly larger number of displaced supercell structures in scaling with supercell size and number of atoms.
We therefore estimated thermal conductivity by employing the 2$\times$2$\times$2 supercells with a series of force calculations for 1280 displaced supercell structures per model.

\section{Results and discussion}
\subsection{Lattice Dynamics and Anharmonicity}
\begin{figure}[!t]
\centering
\includegraphics[clip=true,scale=0.08]{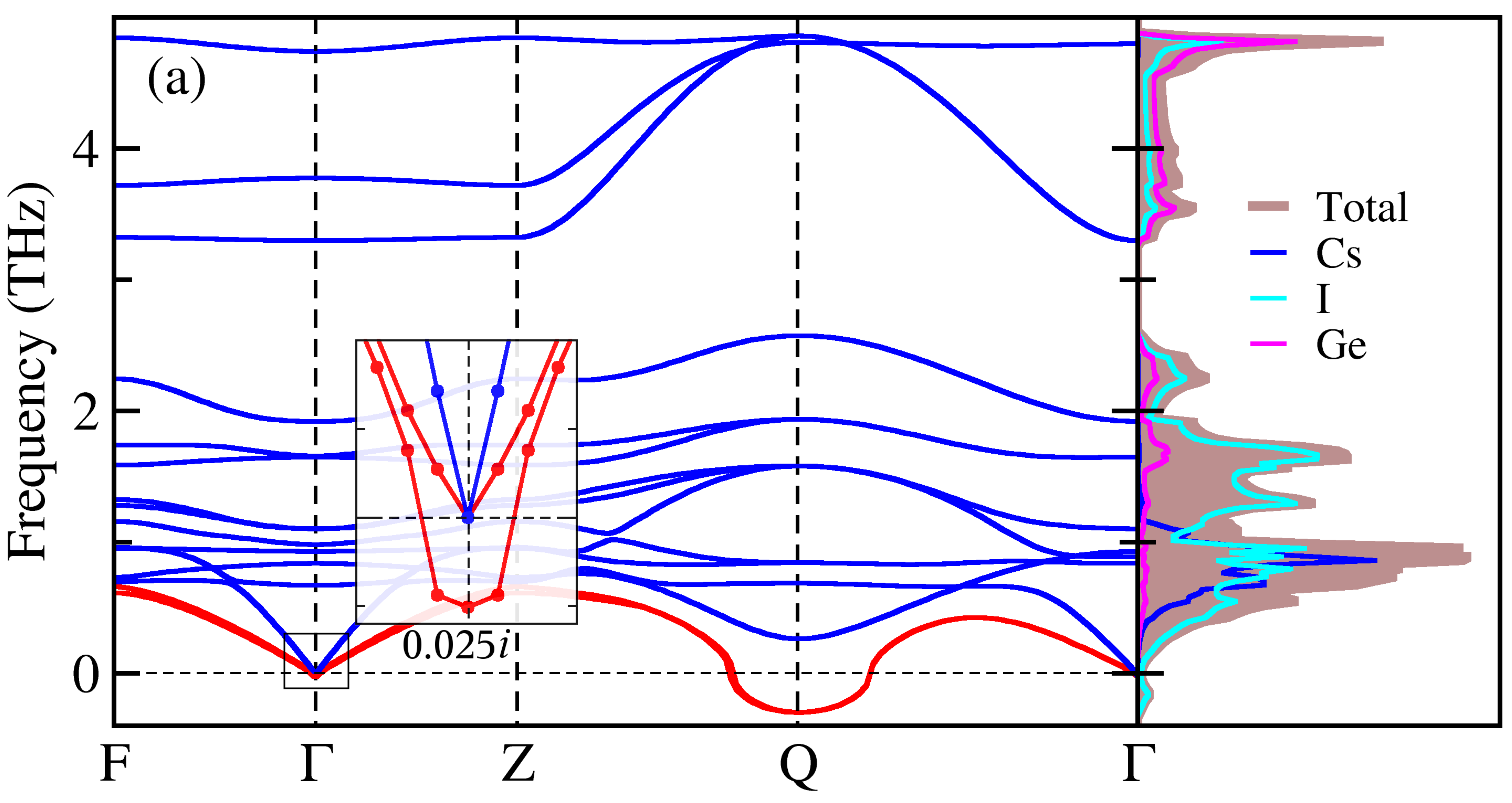}
\includegraphics[clip=true,scale=0.08]{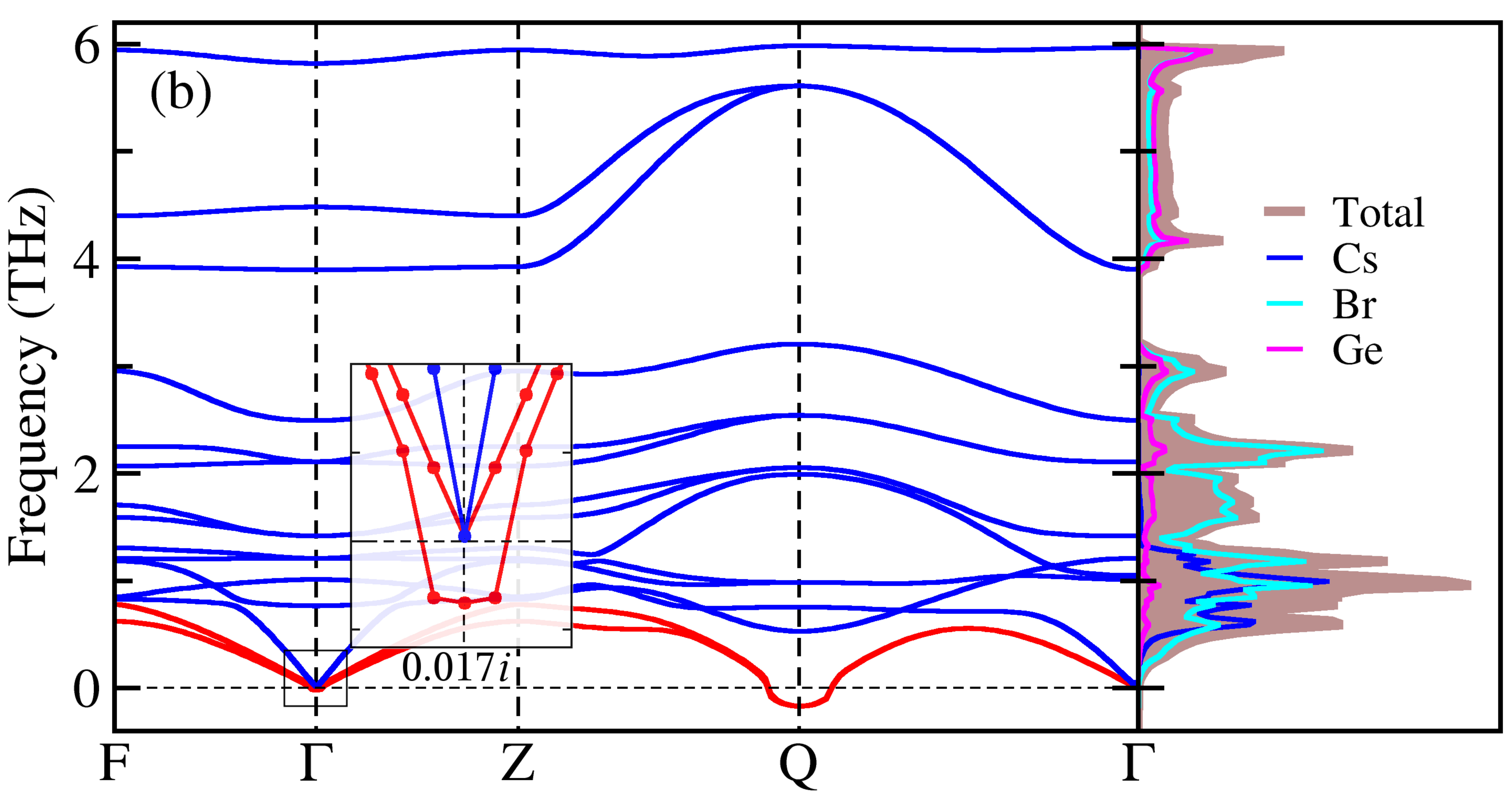}
\includegraphics[clip=true,scale=0.08]{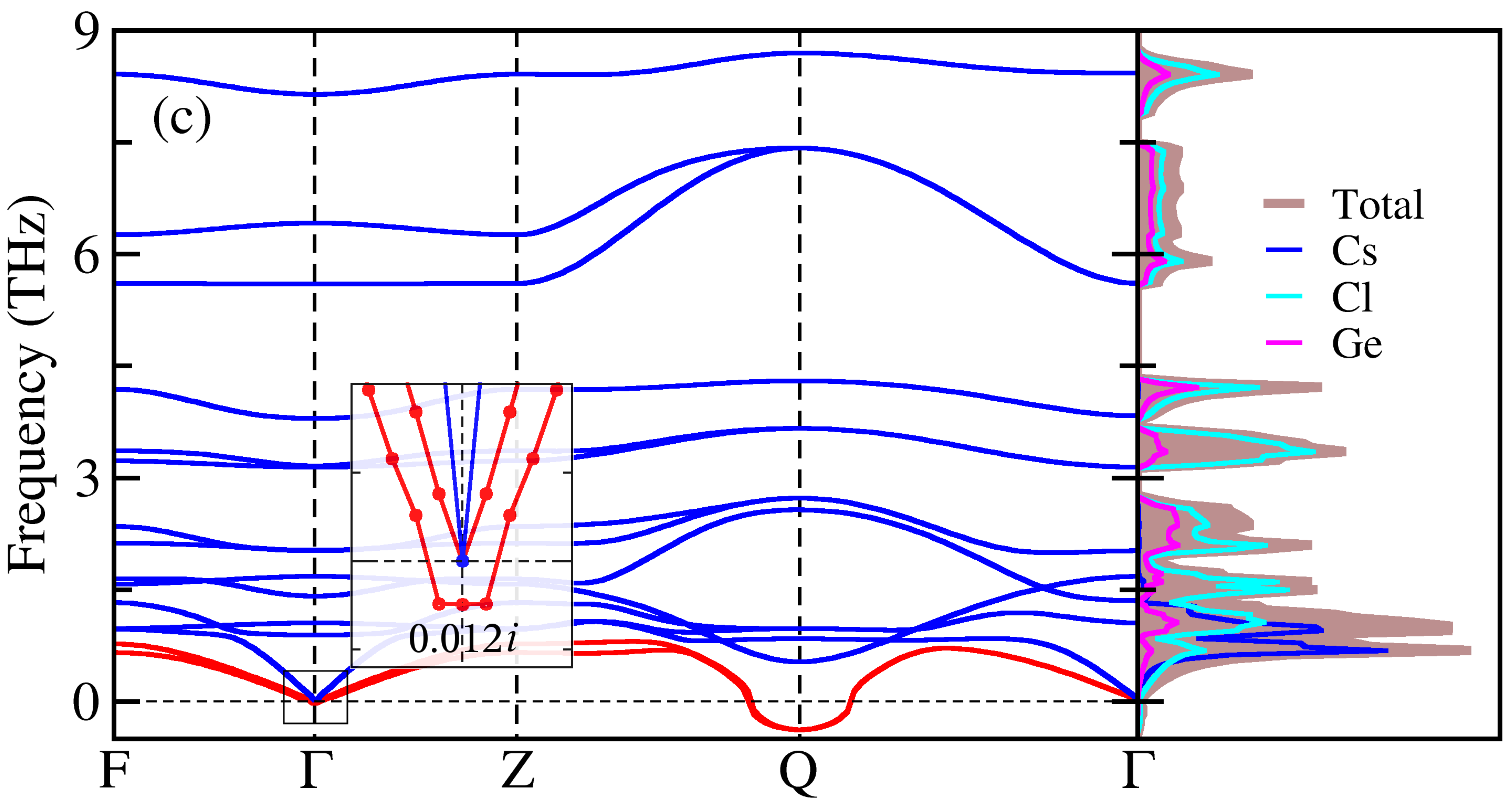}
\caption{Phonon dispersions and atomic resolved phonon density of states of the optimized \ce{CsGeX3} in rhombohedral phase with $R3m$ space group for (a) X = I, (b) X = Br, and (c) X = Cl. The dispersions show anharmonic soft-modes at the $Q$ and $\Gamma$ points of the Brillouin zone, as indicated by red lines. The insets magnifiy the region around the $\Gamma$ point for the soft-modes with the imaginary frequencies of 0.025$i$, 0.017$i$, and 0.012$i$ for X = I, X = Br and X = Cl, repectively.}
\label{fig1}
\end{figure}
Our first-principles lattice dynamics calculations were carried out by applying the finite displacement method as implemented in the {\footnotesize Phonopy} package~\cite{phonopy}.
%Full computational details of our work is found in the Supplemental Material~\cite{supp}.
Figure~\ref{fig1} shows the calculated phonon dispersions of \ce{CsGeX3} (X = I, Br, Cl) in the rhombohedral phase with a $R3m$ space group, along the symmetry line of $F \rightarrow \Gamma \rightarrow Z \rightarrow Q \rightarrow \Gamma$ of the Brillouin zone (BZ), and the atomic resolved phonon density of states (DOS).
Commonly, the anharmonic phonon modes, which have imaginary phonon frequencies, were identified strongly at the BZ boundary point $Q$ and weakly at the zone center point $\Gamma$, demonstrating a dynamic instability of the rhombohedral phase.
At the $Q$ point the anharmonic soft-modes were found to be double-degenerated with the eigenvalues (phonon frequencies) of 0.298$i$, 0.168$i$ and 0.377$i$ THz while at the $\Gamma$ point non-degenerate modes with the eigenvalues of 0.025$i$, 0.017$i$ and 0.012$i$ THz for X = I, Br and Cl, respectively.
Non-analytic correction on phonon eigenvalues was imposed, ensuring the acoustic branches approaching zero and split between longitudinal-optic modes and transverse-optic modes at the $\Gamma$ point.
From observing the atomic-resolved phonon DOS, the anharmonic soft modes are attributed to vibrations of all the constituent atoms.

\begin{figure}[!b]
\centering
\includegraphics[clip=true,scale=0.14]{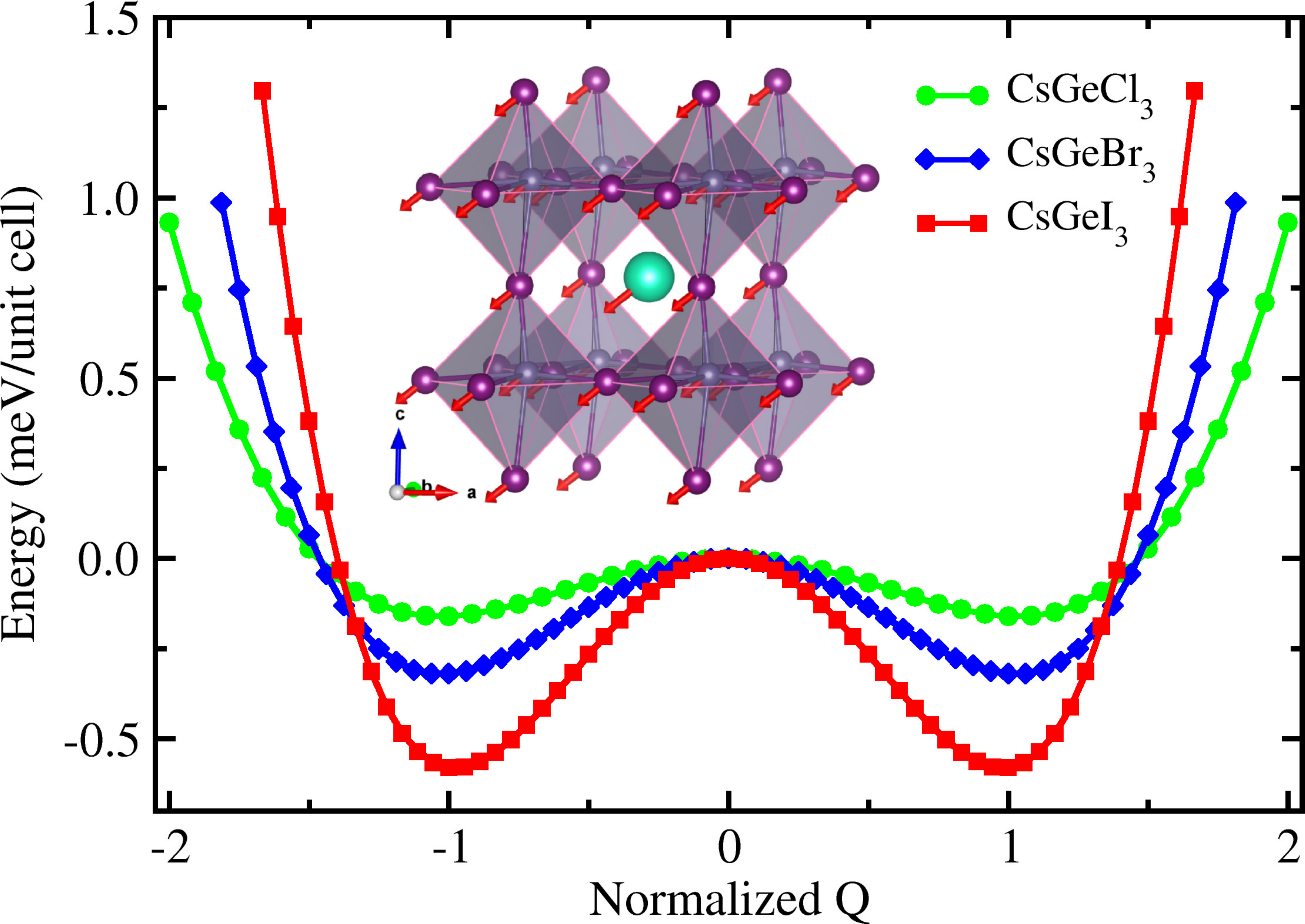}
\caption{Double-well potential as a function of normalized distortion amplitude Q associating with the anharmonic phonon mode at $\Gamma$ point for \ce{CsGeX3}. The inset shows a polyhedral view of unit cell, together with the atomic displacements according to the anharmonic phonon eigenvector. Green- and purple-colored balls represented Cs and X atoms, and gray-colored polyhedron is for \ce{GeX6}.}
\label{fig2}
\end{figure}
\begin{figure*}[!t]
\centering
\includegraphics[clip=true,scale=0.13]{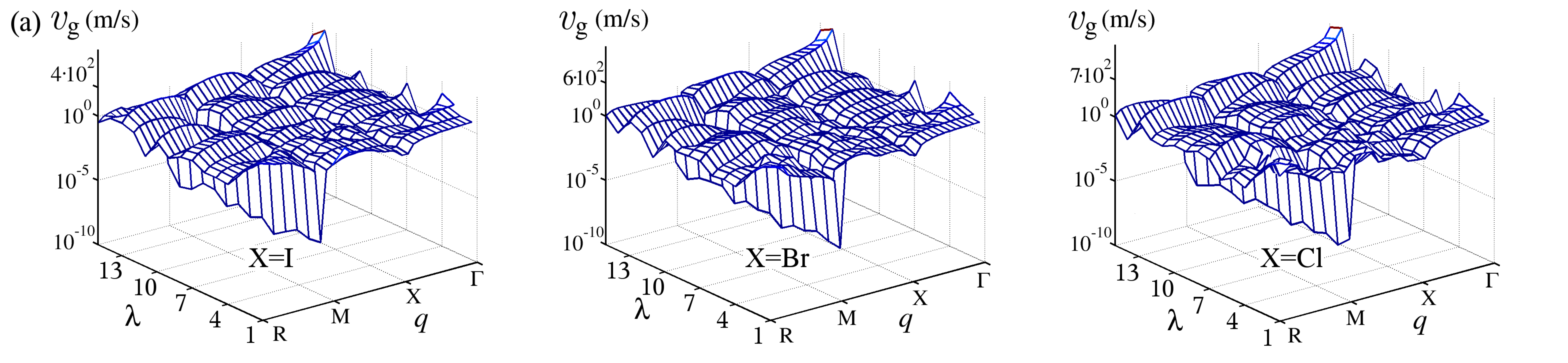} \\
\includegraphics[clip=true,scale=0.13]{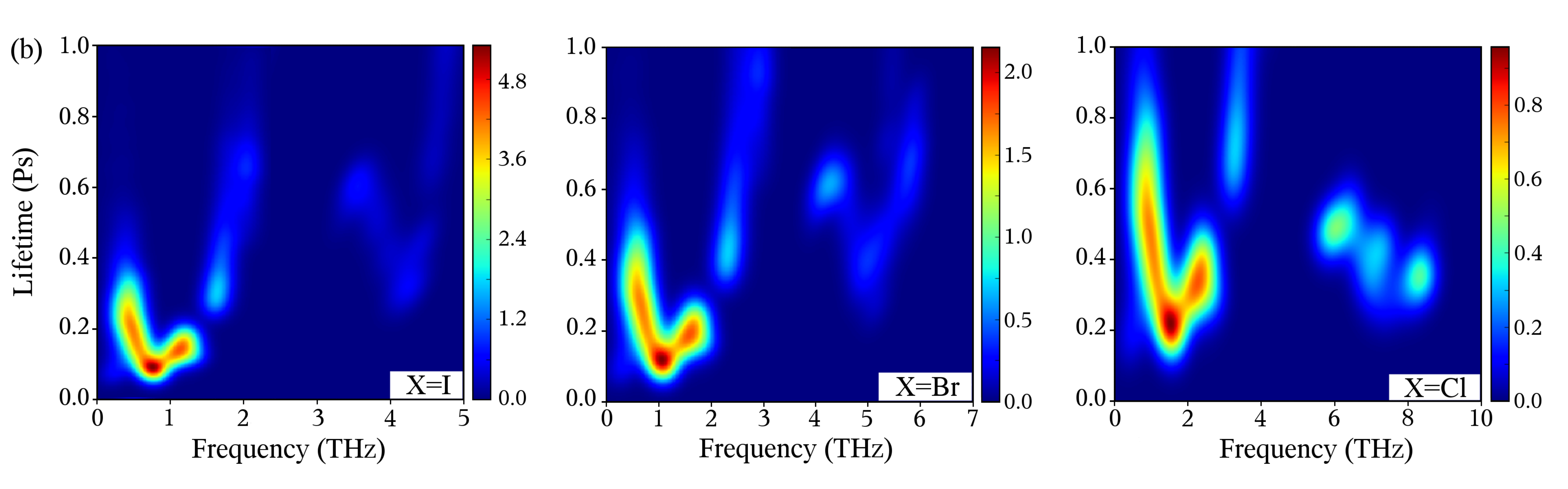} \\
\includegraphics[clip=true,scale=0.13]{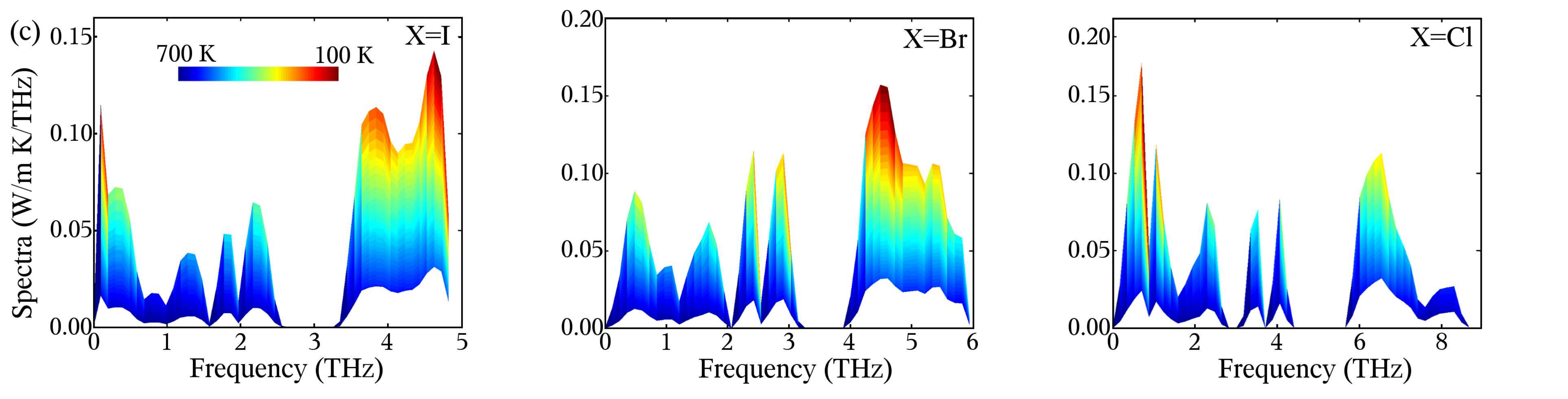} \\
\caption{(a) The calculated group velocity $\textit{\textbf{v}}_{\text{g}}$ as a logarithmic function of phonon momentum $q$ and mode $\lambda$, (b) the phonon lifetime together with phonon DOS in background, calculated at 300 K, and (c) thermal conductivity spectra $\kappa(\omega,T)$ as a function of phonon frequency $\omega$ and temperature $T$ for \ce{CsGeX3} (X = I, Br, Cl).}
\label{fig3}
\end{figure*}
In order to clearly understand the lattice-dynamics related with the soft-modes, we show the atomic displacements along the anharmonic phonon eigenvectors in Fig. S1 (see the Supplemental Material~\cite{supp}).
From the symmetry analysis of the phonon eigenvectors, these soft phonon modes were found to be responsible for symmetry breaking displacive instabilities, being conventional phenomena in both oxide and halide perovskites.
It was revealed that at the $\Gamma$ point the non-degenerate soft mode causes collective displacements of all the constituent atoms on (111) plane, with a larger magnitude of Cs displacement while almost the same magnitudes of X and Ge displacements, resulting in a compression of the Cs$-$X and Cs$-$Ge bonds.
Moreover, the corresponding potential energy surface (PES) shows a well-known double-well shape, as can be seen in Fig.~\ref{fig2}.
The energy was estimated as a function of normalized distortion Q with energy minima at $|Q|=1$.
Depths of the double wells were determined to be 0.58, 0.32 and 0.16 meV per unit cell for X=I, Br and Cl, respectively.
These are smaller in two or three orders of magnitude than those of Pb- or Sn-based halide perovskites ~\cite{Yang17jpcl,Jong19prb,Marronnier17jpcl} and also much smaller than the one of PbTe~\cite{Xia180}, indicating that the anharmonicity at the $\Gamma$ point is not so strong for \ce{CsGeX3}.
On the other hand, the double-degenerated soft modes at the $Q$ point originate complicated \ce{GeX6} octahedral distortions, with much smaller magnitudes of Cs and Ge displacements than X atom, leading to an anti - phase octahedral tiltings, as also observed in the Pb- and Sn-based perovskites~\cite{Yang17jpcl,Jong19prb,Marronnier17jpcl}.
Moreover, when relaxing the structure along the soft-modes at the $\Gamma$ point, the compounds were found to transform to the monoclinic phase with a $Cm$ space group, causing an elimination of soft-modes at the $Q$ point (see Fig. S2 for the phonon dispersion in this phase in the Supplemental Material~\cite{supp}).

\subsection{Lattice Thermal Conductivity}
To get an insight into the thermal transport, we solved the Boltzmann transport equation within the single-mode relaxation-time approximation, as implemented in the {\footnotesize Phono3py} code~\cite{phono3py}.
With this approach, the phonon lifetimes are computed as the phonon self-energy, using the three-phonon interaction strengths obtained from the third-order force constants.
Then, the lattice thermal conductivity can be determined using the formula $\kappa=\Sigma_{q\lambda}C_{\text{v},q\lambda}|\textit{\textbf{v}}_{q\lambda}|^2\tau_{q\lambda}$ with the constant-volume heat capacity $C_{\text{v},q\lambda}$, group velocity $\textit{\textbf{v}}_{q\lambda}$ and relaxation time $\tau_{q\lambda}$, which are dependent on phonon momentum $q$ and mode $\lambda$.
Through the calculation, we found that $C_\textrm{v}$ increases rapidly with increasing temperature from 100 to 300 K and very slowly after that, and enhances clearly going from X = I to Br and to Cl (see Fig. S3 in the Supplemental Material~\cite{supp}).
The calculated group velocity $\textit{\textbf{v}}_{q\lambda}$ as a logarithmic function of $q$ and $\lambda$ is shown in Fig.~\ref{fig3}(a), revealing that its magnitude also enhances decreasing the ionic radius of halogen atom with the maximum velocities of about 400, 600 and 700 m/s$^2$ for X = I, Br and Cl, respectively.
Figure~\ref{fig3}(b) presents the phonon lifetimes $\tau_{q\lambda}$, which clearly get longer upon decreasing the halide ionic radius.
As increasing temperature, meanwhile, they were found to be get shorter (see Fig. S4 for the phonon lifetimes of \ce{CsGeCl3} at different temperatures in the Supplemental Material~\cite{supp}).
For the phonon lifetimes, the three-phonon scattering processes within the lower frequency range are dominative compared to the scattering processes within the upper range.
One common variation tendency of $\tau_{q\lambda}$, $\textit{\textbf{v}}_{q\lambda}$ and $C_{\textrm{v},q\lambda}$ is that they all increase gradually going from X = I to Br to Cl.

\begin{figure}[!t]
\centering
\includegraphics[clip=true,scale=0.47]{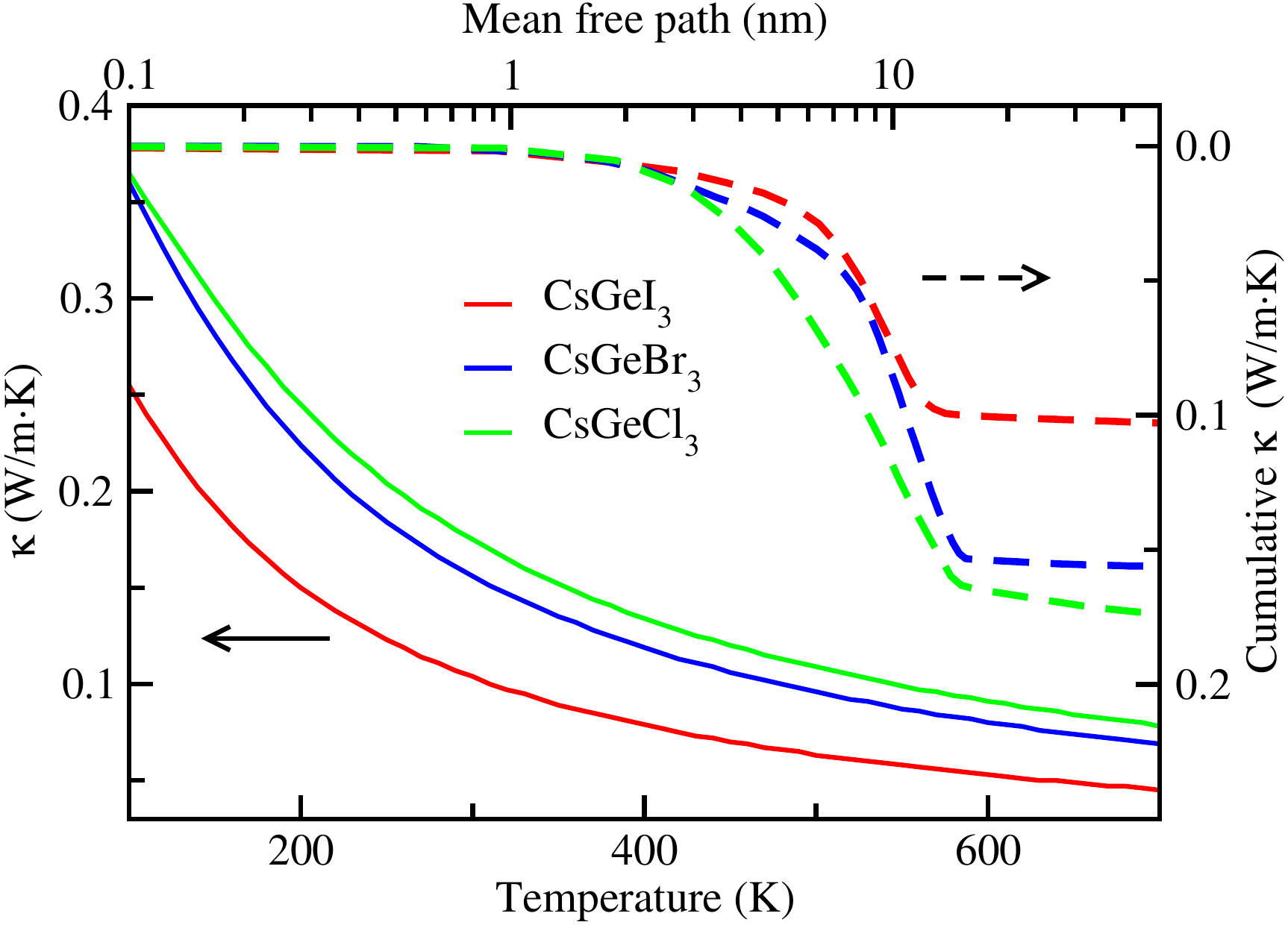}
\caption{The calculated lattice thermal conductivity $\kappa$ (solid lines) as a function of temperature and cumulative $\kappa$ (dashed lines) as a function of mean free path.}
\label{fig4}
\end{figure}
Using the computed $\tau_{q\lambda}$, $\textit{\textbf{v}}_{q\lambda}$ and $C_{\textrm{v},q\lambda}$, we determined the lattice thermal conductivity as a function of temperature.
In the previous work for the strongly anharmonic compounds such as GeTe, PbSe and PbTe, it was found that considering only anharmonic phonon renormalization at finite temperature overestimates the thermal conductivity, whereas only including four-phonon scatterings is responsible for sever reduction of the thermal conductivity ~\cite{Xia180,Xia182,Xia181}. 
In recent work for other type of anharmonic compound \ce{Tl3VSe4}, it has been demonstrated that considering only three-phonon scatterings without anharmonic phonon renormalization may underestimate $\kappa$ in comparison with experiment, whereas considering both the four-phonon scattering and finite temperature effect reproduces the experimental value of $\kappa$, highlighting the importance of high order anharmonicity in the low-$\kappa$ systems such as \ce{Tl3VSe4}~\cite{Xia20}.
That is, by taking into account both the three- and four-phonon scattering processes with temperature-induced anharmonic phonon renormalization, $\kappa$ can be calculated to be in good agreement with the experimental value.
Moreover, Lee \textit{et al.}~\cite{Lee17pnas} found that for $\gamma$-\ce{CsSnI3}, the experimental value $\kappa$ of 0.38 W m$^{-1}$ K$^{-1}$ was in good agreement with the values of 0.33, 0.17 and 0.20 W m$^{-1}$ K$^{-1}$ calculated along the (001), (010) and (100) directions without considering the anharmonic phonon renormalization and four-phonon scatterings.
Such agreement in $\gamma$-\ce{CsSnI3} is attributed to a fortunate error cancellation effect.
However, it should be noted that such error cancellation effect is certainly not universal but may depend strongly on a specific system.
In fact, $\gamma$-\ce{CsSnI3} in the previous work~\cite{Lee17pnas} is in the orthorhombic phase, which is different from our case of \ce{CsGeX3} in the rhombohedral phase.
Based on such considerations, in this work, we considered up to three-phonon interactions while ignoring higher-order phonon scattering processes and anharmonic phonon renormalization at finite temperature for the calculation of thermal transport properties, and we should note that our calculation may underestimate the thermal conductivity due to ignoring those two effects.
By doing convergence test of $\kappa$ with respect to $q$-point mesh, we found the $30\times30\times30$ mesh reliable for calculation of $\kappa$ within a relative error less than 1\% (see Fig. S5 for convergence test in the Supplemental Material~\cite{supp}).

Figure~\ref{fig4} shows the calculated lattice thermal conductivity $\kappa$ as a function of temperature.
As such, $\kappa$ was found to gradually decrease when increasing temperature, and to increase going from X = I to Br to Cl at the given temperature.
We depicted the variation tendency of the thermal conductivity spectra $\kappa(\omega,T)$ as a function of phonon frequency $\omega$ and temperature $T$ in Fig.~\ref{fig3}(c), giving an evidence of that $\kappa$ decreases with the increment of temperature.
From the cumulative $\kappa$ curves in Fig.~\ref{fig4}, we revealed that the maximum length of mean free path is about 13 nm and the major heat-carrying phonon modes have the lengths from 1 to 10 nm.
At room temperature of 300 K, the ultralow thermal conductivities were obtained as 0.10, 0.16 and 0.18 W m$^{-1}$ K$^{-1}$ for X = I, Br and Cl, respectively.
When compared to 1.7 and 2.5 W m$^{-1}$ K$^{-1}$ of GeTe~\cite{Wu14} and PbTe~\cite{Bagieva12}, these are much smaller in one order of magnitude, implying that \ce{CsGeX3} can have a high value of $ZT$.

\subsection{Carrier Mobility}
As a final step, we estimated the carrier mobility by employing the deformation potential theory (DPT)~\cite{Zasavitskii,DPT,Bruzzone}, which can give an important insight into the electrical conductivity.
Within DPT, the carrier mobility is computed using the formula $\mu=\frac{(8\pi)^{1/2}\hbar^4eC}{3(m^*)^{5/2}(k_BT)^{3/2}D^2}$, where $C$, $D$, and $m^*$ are the elastic constant, deformation potential, and effective mass of carrier, respectively.
Among them, the deformation potential, which represents the coupling between electronic bands and lattice vibrations especially corresponding to the acoustic phonon modes, can be obtained by computing the shift of conduction band minimum (CBM) via dilating and compressing the volume of unit cell (see Fig. S6 for the CBM change in the Supplemental Material~\cite{supp}).

In order to thoroughly investigate the electronic transport properties, the electron-phonon interactions should be fully considered by taking into account the coupling with not only the acoustic but also the polar optical phonons.
Recently, Ponce \textit{et al.}~\cite{Ponce19} applied the state-of-the-art schemes based on many-body first-principles calculations to estimate the carrier mobility of \ce{MAPbI3} and \ce{CsPbI3}, demonstrating that the low-energy longitudinal-optical phonons associated with Pb-I bonds play a critical role in suppressing mobility at room temperature. 
By applying the DPT and DFT + SOC method, meanwhile, Ying \textit{et al.}~\cite{Ying18} obtained the electron carrier mobility for \ce{CsSnI3} in the cubic phase to be 400 cm$^2$ V$^{-1}$ s$^{-1}$, being comparable with the experimental value of 394 cm$^2$ V$^{-1}$ s$^{-1}$ in the orthorhombic phase~\cite{Lee17pnas}.
It should be noted that as compared to the many-body first-principles calculations, the carrier mobility might be overestimated within DPT in this work because of ignoring the coupling between electronic bands and lattice vibrations associated with the optical phonon modes.

\begin{table}[!t]
%\small
\caption{The calculated effective mass $m^*$, elastic constant $C$ (GPa), deformation potential $D$ (eV), carrier mobility $\mu$ (cm$^2$ V$^{-1}$ s$^{-1}$) and thermal conductivity $\kappa$ (W m$^{-1}$ K$^{-1}$) at 300 K in \ce{CsGeX3} (X = I, Br, Cl). For comparison, the experimental values of the conventional thermoelectrics, PbSe and PbTe~\cite{Wang12, Bagieva12} are also shown.}
\label{tab1}
\begin{tabular}{lccccc}
\hline
      & $m^*$ & $C$ & $D$ & $\mu$ & $\kappa$ \\
\hline     
\ce{CsGeI3}   & 0.22 & 46.17 & 10.81 & 1677 & 0.10\\
\ce{CsGeBr3}  & 0.27 & 56.62 & 10.14 & 1401 & 0.16\\
\ce{CsGeCl3}  & 0.36 & 63.28 &  9.53 & 863 & 0.18\\
PbSe          & 0.27 & 71.00 &25.00 & 1140 & 1.91\\
PbTe          & 0.26 & 91.00 &22.00 & 1508 & 2.50\\
\hline
\end{tabular}
\end{table}
Table~\ref{tab1} lists the calculated values for these quantities and the thermal conductivity at 300 K for \ce{CsGeX3}, with the experimental values of PbSe and PbTe~\cite{Wang12, Bagieva12}.
At 300 K, the mobilities were found to be 1677, 1401 and 863 cm$^2$ V$^{-1}$ s$^{-1}$ for X=I, Br and Cl, respectively.
These are comparable to the values of 1140 and 1508 cm$^2$ V$^{-1}$ s$^{-1}$ of PbSe and PbTe, being attributed to the joint effects of low deformation potentials and small effective masses.
The ultralow thermal conductivities and high mobilities predicted in \ce{CsGeX3} indicate their great potential of high performance thermoelectrics.

\section{Conclusions}
In conclusion, we have introduced a new class of thermoelectrics, Ge-based halide perovskites \ce{CsGeX3} (X = I, Br, Cl), and we have quantified their anharmonic lattice dynamics, ultralow thermal conductivities and high carrier mobilities by applying first-principles methods.
These inorganic halide perovskites have a variety of attractive and potentially profitable properties for photovoltaic and thermoelectric devices.
Furthermore, this work highlights the way of looking beyond well-studied perovskites in the search for thermoelectrics with high performance and low cost.

\section*{Acknowledgments}
This work was supported by Grants No. 2016-20 from the State Commission of Science and Technology, DPR Korea.
Computations were performed on the HP Blade System C7000 (HP BL460c) that is managed by Faculty of Materials Science, Kim Il Sung University.

\section*{\label{note}Notes}
The authors declare no competing financial interest.

\bibliographystyle{elsarticle-num-names}
\bibliography{Reference}

\end{document}